\documentclass[12pt,letter]{article}
\usepackage{amsmath}
\usepackage{authblk}
\usepackage[group-separator={,}]{siunitx}
\usepackage{booktabs} 
\usepackage{graphicx,psfrag,epsf}
\usepackage{enumerate}
\usepackage{bbm}
\usepackage{url} 

\usepackage{amssymb}
\usepackage{comment}
\usepackage{caption}
\usepackage{subcaption}
\usepackage{mathtools}
\mathtoolsset{showonlyrefs}
\usepackage{pgfplots}
\usepackage{filecontents}
\usepackage{kpfonts}
\usepackage{float}
\usepackage{amsthm}

\usepackage{graphicx,subcaption}
\usepackage{multirow}
\usepackage{fullpage}
\usepackage{tikz}
\usepackage{pgfplotstable}
\usepgfplotslibrary{groupplots}
\usepgfplotslibrary{polar}
\usepgfplotslibrary{external}
\definecolor{myblue}{rgb}{.5, .5, 1}

\usetikzlibrary{decorations.text,shadows}
\usepackage{ifthen}
\usetikzlibrary{matrix,positioning}
\usetikzlibrary{shapes.misc}
\usetikzlibrary{plotmarks}

\tikzset{cross/.style={cross out, draw=black, minimum size=2*(#1-\pgflinewidth), inner sep=0pt, outer sep=0pt},
cross/.default={12pt}}
\usepackage{setspace}

\date{April 2017}

\begin{document}

\title{\bf Detecting periodic subsequences in cyber security data}
\author[1]{Matthew Price-Williams}
\author[1,2]{Nick Heard}
\author[3]{Melissa Turcotte}
\affil[1]{Department of Mathematics, Imperial College London}
\affil[2]{Heilbronn Institute for Mathematical Research, University of Bristol}
\affil[3]{Advanced Research in Cyber Systems, Los Alamos National Laboratory}
\maketitle
\bigskip
\begin{abstract}
Statistical approaches to cyber-security involve building realistic probability models of computer network data. In a data pre-processing phase, separating automated events from those caused by human activity should improve statistical model building and enhance anomaly detection capabilities. This article presents a changepoint detection framework for identifying periodic subsequences of event times. The opening event of each subsequence can be interpreted as a human action which then generates an automated, periodic process. Difficulties arising from the presence of duplicate and missing data are addressed.
The methodology is demonstrated using authentication data from the computer network of Los Alamos National Laboratory.
\end{abstract}
\section{Introduction}

Recent statistical approaches to cyber-security defence \cite{lazarevic,neil} utilize anomaly detection techniques based upon statistical models of normal network behavior, analyzing deviations in an attempt to identify malicious actors. 
Regarding the arrivals of communications between each pair of hosts in a computer network as a point process of event times on $\mathbb{R}^+$, this article proposes a method to distinguish between automated network events and those caused by human behavior.

Polling behavior at a constant periodicity is a common feature of automated signal traffic. \cite{polling} presents a method to detect overall polling behavior in a sequence of event data using Fourier analysis. The work presented here aims to detect more complex polling behavior where the entire traffic sequence can be split up into periodic subsequences, separated by more random durations of inactivity. In this scenario, the start of each periodic subsequence is a user-driven event that initiates subsequent automated polling events that beacon at a constant periodicity.
Since the user-driven events are not labeled, inferring their identities between the bulk of automated, periodic events can be viewed as a changepoint detection problem. 

When the method of \cite{polling} is applied to periodic subsequences of event data, detection performance can be seen to deteriorate when the lengths of each periodic subsequence are relatively short. This provides justification for devising bespoke procedures for detecting periodic subsequences.  

Treating automated and user-driven data separately should provide a more robust framework for modeling network data, whereby bespoke models can be specified for each type of behavior. Alternatively, as a data reduction tool, identifying only user-driven events represents a significant thinning of the bulk of computer network data, potentially improving anomaly detection capabilities. Examples of periodic subsequences in network data can occur in the constant refreshing of an open webpage, or when validating log on credentials in authentication data \cite{kent1}. 

This article focuses on two types of polling behavior which can be observed in computer network data, each exhibiting a hypothetical periodicity $P$ in different ways:
\begin{enumerate}
\item Fixed phase polling: Event times occur every $P$ seconds plus a random zero-mean error; any delay in one event time does not propagate into future event times.
\item Fixed duration polling: Event times occur $P$ seconds after the preceding event, plus a random zero-mean error.
\end{enumerate}
In this article a full generative model is proposed for event times exhibiting both types of polling. These models are robust to both missing and duplicate data. 

Section \ref{sec:CN} will describe the data used for analysis, taken from the  Los Alamos National Laboratory computer network. 
Section \ref{sec:DPB} will introduce the screening method used for detecting periodicity in a sequence of event times, demonstrating effectiveness in the presence of periodic subsequences. Section \ref{sec:DS} will show how directional statistics can be used to model the error terms of periodic events as angular displacements. Sections \ref{sec:M} and \ref{sec:CD} will describe how changepoint detection methods can identify separate periodic subsequences conditional on an underlying model. Section \ref{sec:7} will present results from both a simulated data example which exhibits fixed phase polling, and the real computer network data which exhibits fixed duration polling.

\section{Computer Network Authentication}
\label{sec:CN}
The data analyzed are authentication logs collected over 58 days from the Los Alamos National Laboratory (LANL) internal computer network \cite{kent2,kent1}.
The authentication logs record a source username and computer, a destination username and computer, and the time when the authentication event was initiated. Additionally, the type of authentication event is also recorded, such as ``Network'', ``Kerberos'' or ``Negotiate''. 
Authentication events often exhibit periodic polling, separated by random durations of inactivity. This can occur when a user initially authenticates their credentials on a computer, and the authentication mechanism then periodically verifies the continued validity of those credentials. Such information cannot be derived from the authentication event types collected in the data.

\begin{figure}[!ht]
\begin{minipage}[c]{0.5\textwidth}
  \centering
  \includegraphics[width=\textwidth]{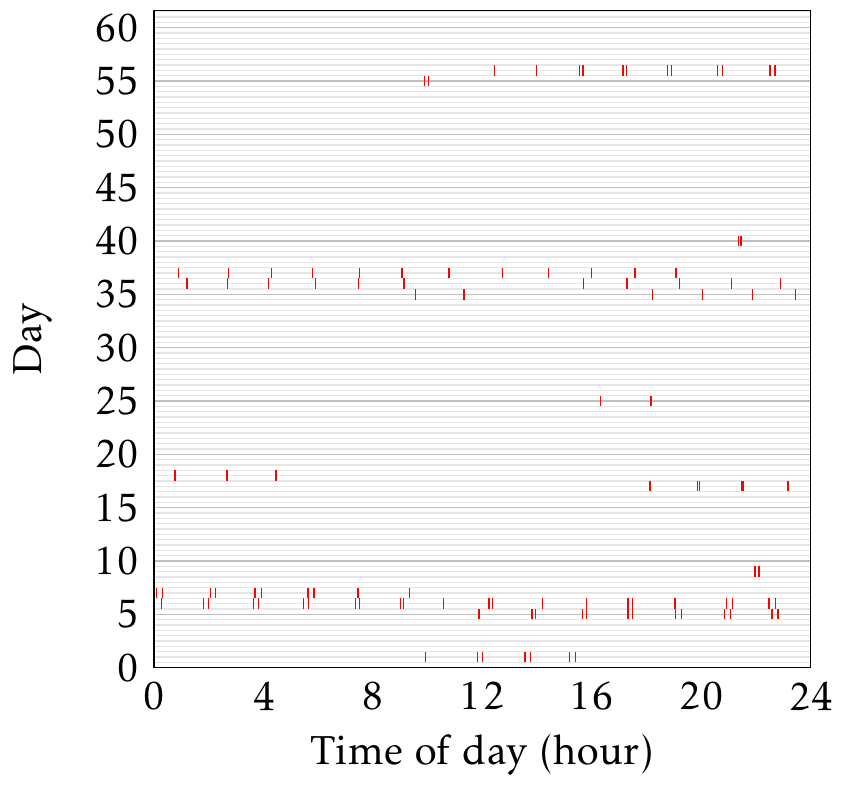}

\end{minipage}
\hfill
\begin{minipage}[c]{0.5\textwidth}
  \centering
  \includegraphics[width=\textwidth]{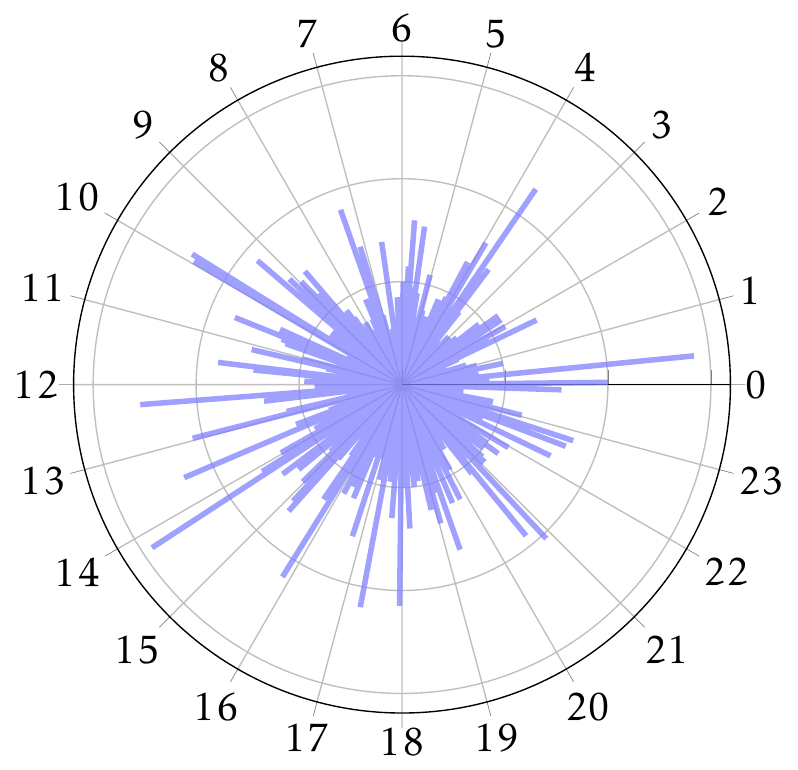}
\end{minipage}
\caption{Log on event times over a 58-day period for an example user, U514, from the LANL computer network. Left: Log on event times between two specific computers frequented by the user, C$528$ and C$15607$. Right: The time of day distribution for all events generated by the user.}
\label{fig:logon}
\end{figure}

The left panel of Figure \ref{fig:logon} plots all event times for an example user, U514, authenticating between two particular computers, C528 and  C15607, over the 58-day period. It is apparent that these data comprise several periodic subsequences of evenly-spaced event times. Each subsequence starts during working hours, and so these initial events are potentially user-driven. Note that within the periodic subsequences there can be duplicate events or missing data, and so it is important that any proposed model is robust to these data features. Duplicates can occur when multiple packets are sent, and missing data can occur when there is packet loss either on the network or on the central authentication server. A comprehensive explanation of the data is given in \cite{kent1}.

The right panel of Figure \ref{fig:logon} shows a circular histogram of the times of day for all \num{4688} ``Log On'' events initiated by the same user, U514, connecting between any source and destination computers. Time-of-day event data are most informatively plotted on a circle whose circumference represents one day. In this way, event data falling just before and just after midnight are displayed close together, rather than at opposite ends of a section of the real line.
The limited variation in the number of events within and outside the working day further implies that there is substantial automated polling behavior.

\section{Detecting Polling Behavior} 
\label{sec:DPB}

A natural approach for detecting polling behavior, described in detail in \cite{polling}, is to calculate a discrete Fourier transform \cite{halliday} of the event times of each point process in the network in order to uncover any periodicities; a large peak in the resulting periodogram $S(f)$ indicates polling behavior at frequency $f$. In particular, a standard significance test for a simple periodicity (see, for example \cite{jenkins}) uses the test statistic
\begin{equation}
g=\max_f \frac{S(f)}{\sum_{f'} S(f')},\label{eq:g_test}
\end{equation}
where the sum in the denominator and the maximization are both over the Fourier frequencies. An approximate upper-tail p-value for $g$ can be calculated under the null hypothesis of no polling behavior.

\cite{turcotte15} presents a Bayesian model for network authentication behavior which was applied to the data described in Section \ref{sec:CN} with some success. As a pre-processing step, any edges within the network that were identified as containing significant polling behavior according to the $g$-statistic procedure of \cite{polling} were removed from the data. This level of filtering could lead to potentially important user-driven authentication behavior also being deleted. The proposed methodology seeks to separate the user-driven events from the automated polling events on a given network edge, rather than completely removing the edge.

\subsection{Detecting Polling Subsequences}
\label{sec:DPS2}

This section presents a simulation study to demonstrate the performance of the Fourier analysis presented in Section \ref{sec:DPB} for detecting polling behavior contained within periodic subsequences. This study provides justification for using a changepoint detection methodology to detect periodic subsequences.  
To construct a full generative model for an ordered sequence of event times with intermittent periodicity, we first consider two sequences
\begin{equation}
\label{setup1}
\begin{split}
x_1,x_2,\ldots, x_S &\overset{i.i.d.}\sim \text{Exponential}(\lambda),\\
n_1,n_2,\ldots, n_S &\overset{i.i.d.}\sim \text{Geometric}(q),
\end{split}
\end{equation}
such that $x_i$ specifies the duration of inactivity before the $i$th polling subsequence commences, $n_i$ specifies the number of beaconing periods of length $P$ within that $i$th subsequence and $S$ is the number of polling subsequence within the event sequence. 

In each simulation a sequence is sampled from the model \eqref{setup1} for different values of $q$ and $S$. The constant periodicity is fixed at $P=1$ and $\lambda$ is chosen to be $0.2$. The value for $\lambda$ generates large durations of inactivity which break the polling cycle. The event times in the beaconing subsequences are perturbed with U$(-0.2,0.2)$ errors;
more realistic representations of this error are provided in Section \ref{sec:DS}.

The case where $q=1$ corresponds to no periodic behavior, and the approximate $p$-values from the upper tail of the test statistic \eqref{eq:g_test} will be approximately uniformly distributed on $[0,1]$. As $q$ decreases, the proportion of periodic data increases and the test should yield statistically smaller $p$-values. For $q<1$, increasing the total number of subsequences $S$ increases the overall sample size, and should also yield statistically smaller $p$-values.

\begin{figure}[th!]
  \centering
  \includegraphics[width=\textwidth]{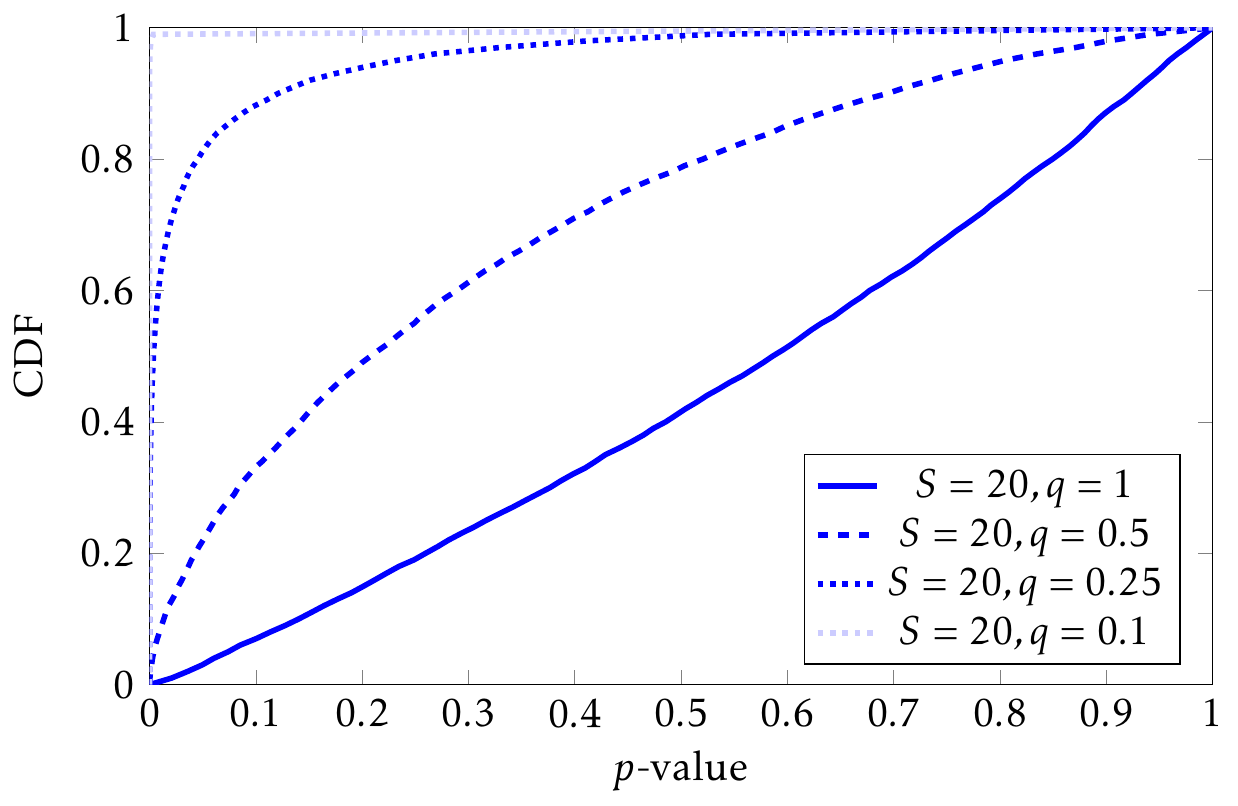}
\caption{Distribution of $p$-values from $g$-test of simple periodicity for $20$ polling subsequences generated from \eqref{setup1} with parameters $\lambda=0.2$ and $q\in\{1,0.5,0.25,0.1\}$. It is clear that a decrease in $q$ yields statistically smaller $p$-values for periodicity in the event sequence.}  
\label{fig:1}
\end{figure}

\begin{figure}[th!]
  \centering
  \includegraphics[width=\textwidth]{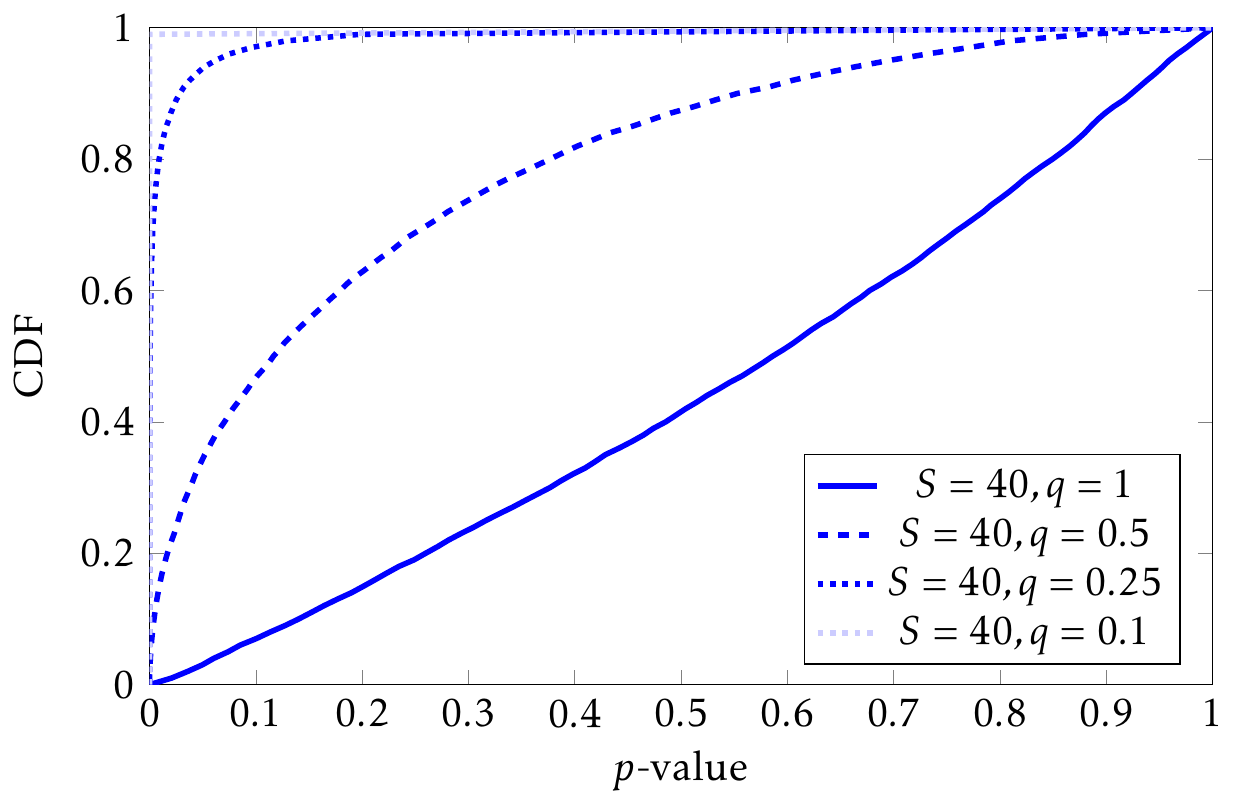}
\caption{Distribution of $p$-values from $g$-test of simple periodicity for $40$ polling subsequences generated from \eqref{setup1} with parameters $\lambda=0.2$ and $q\in\{1,0.5,0.25,0.1\}$. By comparing the results with Figure \ref{fig:1} it is clear that increasing the number of polling subsequences yields statistically smaller $p$-values for $q\in\{0.5,0.25,0.1\}$.}
\label{fig:1a}
\end{figure}

Ten thousand Monte Carlo simulations were performed for different combinations of $q\in\{1,0.5,0.25,0.1\}$, $S\in\{20,40\}$.   
Figures \ref{fig:1} and \ref{fig:1a} show the empirical cumulative distribution function of the resulting $p$-values for all combinations of $q$ and $S$.

When $q=1$, the $p$-values are approximately uniformly distributed. Even when $q=0.5$, which in expectation provides the shortest meaningful subsequence length for exhibiting a common periodicity, the distribution of $p$-values is already quite different. For a moderate expected subsequence length of $10$, $q=0.1$, the $p$-values from observing $20$ such subsequences are concentrated strongly at zero. These simulations imply that the random lengths of inactivity in-between periodic subsequences do not significantly reduce the signal of the Fourier transform-based $g$-test. Polling behavior is still detectable even in the most extreme cases.
By comparing Figures \ref{fig:1} and \ref{fig:1a}, it is clear that increasing the number of polling subsequences causes a further decrease in the p-values.

This section uses a crude uniform estimate to model the random error of each event in a periodic subsequence. Section \ref{sec:DS} introduces directional statistics and explains how they can be used to provide a smoother more accurate estimate of this error.   

\section{Directional Statistics}
\label{sec:DS}
\begin{figure}[!ht]
    \pgfplotstableread[header=false]{synthetic_events.txt}\SE
    \pgfplotstablegetrowsof{\SE}
    \pgfmathsetmacro\nse{\pgfplotsretval-1}
\pgfplotscreateplotcyclelist{my cycle}{mark=x,mark=square,mark=triangle,mark=o,mark=star}
\begin{minipage}[c]{0.6\textwidth}
  \centering
  \includegraphics[width=\textwidth]{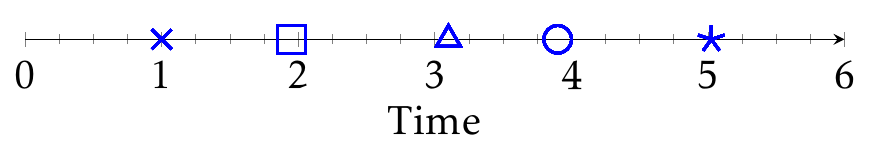}
\end{minipage}
\begin{minipage}[c]{0.4\textwidth}
  \centering
  \includegraphics[width=\textwidth]{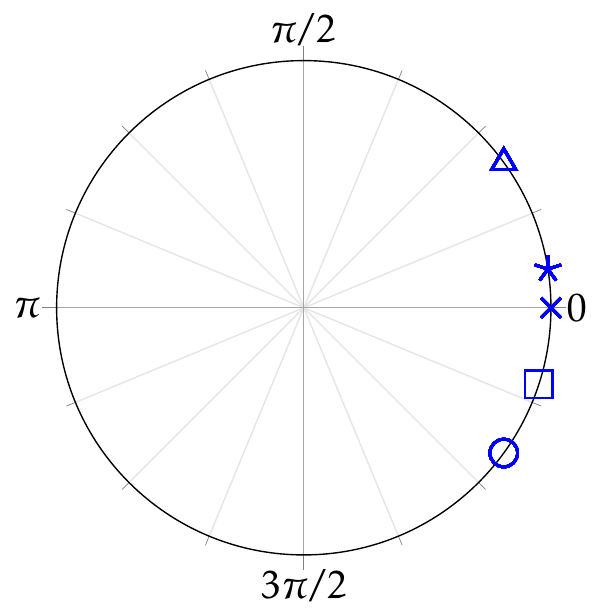}
\end{minipage}
\caption{Cartoon depiction of the transformation $y\mapsto \phi(y)$ reducing event times on $\mathbb{R}^+$ to angular positions with respect to a unit periodicity.}
\label{fig:transformation}
\end{figure}

For modeling data with an underlying periodicity, it is intuitively most simple to consider noise in the data as angular displacements from the underlying periodic sequence. For a given period $P$, event times from a point process can be transformed to directions in two-dimensional space, represented by points on a unit circle. For an event time $y$, let
\begin{align}
\phi(y)=\dfrac{2\pi y}{P}\pmod{2\pi},
\label{eq:angular}
\end{align}
denote the angular position of the corresponding point on the unit circle. Under this transformation, any variability in the periodic event times corresponds to small angular displacements from an overall angular mean. This transformation is depicted in Figure \ref{fig:transformation}.

The mean direction and circular variance provide respective measures of the location and spread of directional data: For a given sample $ \phi= (\phi_1,\ldots,\phi_n)$, let $x_i=(x_{i,1},x_{i,2})$ where $x_{i,1}=\cos\phi_i$ and $x_{i,2}=\sin\phi_i$, such that $x_i$ is the location of a unit vector with angle $\phi_i$ in two-dimensional space. Then let $\bar{x}=(\bar{x}_1,\bar{x}_2)$ denote the mean resultant vector of the sample, where $\bar{x}_1=\sum_{i=1}^nx_{i,1}/n$ and $\bar{x}_2=\sum_{i=1}^nx_{i,2}/n$, and let $\bar{R}=\sqrt{\bar{x}_1^2+\bar{x}_2^2}$ denote the length of $\bar{x}.$
The mean direction $\bar{\phi}$ is defined to be
\begin{equation}
\bar{\phi}=\arctan(\bar{x}_2/\bar{x}_1)\pmod{2\pi},
\end{equation}
and the circular variance is defined as $V=1-\bar{R}$.
A comprehensive overview of these and other directional data summaries is provided in \cite{mardia}.

\subsection{The von Mises Distribution}
A commonly used distribution for directional data is the \textit{von Mises} distribution \cite{vM}. A random variable $\theta$ is said to follow the von Mises distribution with location parameter $\nu\in[0,2\pi)$ and precision parameter $\kappa>0$, written $M(\nu,\kappa)$, if it has density
\begin{align}
\label{eq:13}
f_M(\theta\mid \nu,\kappa) = \frac{\exp(\kappa\cos(\theta-\nu))}{2\pi I_0(\kappa)}, \qquad \theta \in [0,2\pi),
\end{align}
where $I_\ell(\cdot)$ is the modified Bessel function of order $\ell$. In Section \ref{sec:M} the von Mises distribution will be used to construct a model for periodic event subsequences, and so it is useful to review frequentist and Bayesian parameter estimation for this distribution.

\subsubsection{Maximum Likelihood Estimation}\label{sec:mle}
Let $\theta_1,\ldots,\theta_n$ be a sequence of realizations drawn from $M(\nu,\kappa)$. The 
log-likelihood function for this sample simplifies to
\begin{equation}
\label{eq:14}
l(\nu,\kappa\mid\theta_1,\ldots,\theta_n)=n\{\log2\pi+\kappa \bar{R}\cos(\bar{\theta}-\nu)-\log{I_0(\kappa)}\},
\end{equation}
\cite{mardia}. Maximizing with respect to the location parameter $\nu$ yields the maximum likelihood estimate (MLE)
\begin{equation}
\label{nu1}
\hat{\nu}=\bar{\theta},
\end{equation}
where $\bar{\theta}$ is the mean direction of the sample, defined above.
Substituting $\bar{\theta}$ into \eqref{eq:14} and differentiating with respect to $\kappa$ yields the equation 
\begin{equation}
\label{eq:15}
\hat{\kappa}=A^{-1}(\bar{R})
\end{equation}
for the MLE for $\kappa$, where
\begin{equation}
  A(\kappa)=\frac{I_1(\kappa)}{I_0(\kappa)}.
\end{equation}
There is no analytic solution for \eqref{eq:15}, but numerical estimates can be obtained \cite{banerjee,mardia}.

\subsubsection{Bayesian Inference}
\label{BA}
For relatively straightforward Bayesian inference for the von Mises distribution, 
\cite{guttorp} and \cite{damien} use the conjugate prior
\begin{equation}
\label{VMprior}
g(\nu,\kappa)\propto \{I_0(\kappa)\}^{-c}\exp\left(\kappa R_0\cos(\nu-\nu_0)\right).
\end{equation}
This specification is analogous to having observed $c$ notional prior directional samples with a mean direction $\nu_0$ and resultant length $R_0$.

As in the previous section, let $\theta_1,\ldots,\theta_n$ be samples drawn from $M(\nu,\kappa)$. Then the posterior distribution is given by 
\begin{align}
\label{VMpos}
g(\nu,\kappa\mid\theta_1,\ldots,\theta_n)\propto \{I_0(\kappa)\}^{-(c+n)}\exp\left(\kappa R_n\cos(\nu-\nu_n)\right),
\end{align}
where $\nu_n$ is the mean direction of the resultant of the sum of a vector with direction $\nu_0$ and length $R_0$ together with unit vectors in each of the directions of the samples $\theta_1,\ldots,\theta_n$, and $R_n$ is the magnitude of that resultant vector.
The Metropolis Hastings algorithm can be used to obtain a posterior estimate for the precision parameter $\kappa$.

\section{A Model for Periodic Subsequences}
\label{sec:M}
To construct a full generative model for an ordered sequence of event times with intermittent periodicity, we first consider the two sequences defined in \eqref{setup1} on page \pageref{setup1}, where $x_i$ specifies the duration of inactivity before the $i$th polling subsequence commences and $n_i$ specifies the number of beaconing periods of length $P$ within that $i$th subsequence. 

Within a polling subsequence, 0 events (in the case of missing data), 1 event, or multiple events (in the case of duplication) may be observed during each period. Within the $i$th polling subsequence, for $j=1,\ldots,n_i$ let $m_{i,j}$ be the number of events observed in the $j^{th}$ period. The values of each event count $m_{i,j}$ is assumed to be \textit{hurdle geometric}: Defining
\begin{equation}
  \delta_{i,j}=\mathbbm{1}(m_{i,j}>0)
  \label{eq:delta_ij}
\end{equation}
we assume $\delta_{i,j}\sim\mbox{Bernoulli}(1-p)$; if $\delta_{i,j}=0$ then clearly $m_{i,j}=0$, and otherwise if $\delta_{i,j}=1$ then $(m_{i,j}-1)\sim\mbox{Geometric}(r)$ for parameters $0<p,r<1$, implying
\begin{equation}
  \mathbb{P}(m_{i,j}=m)=
  \begin{cases}
    p, & m=0,\\
    (1-p)(1-r)r^{m-1}, & m\geq 1.
  \end{cases}
\label{eq:r}
\end{equation}

\subsection{Fixed Phase Polling}
\begin{figure}[th]
  \centering
\includegraphics[width= \textwidth]{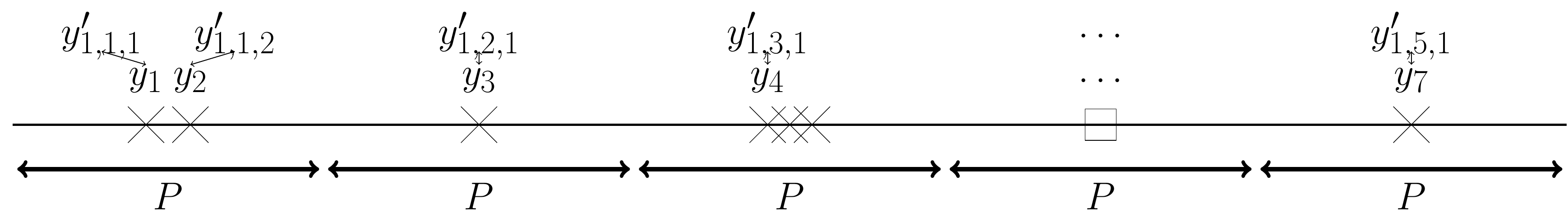} 
\caption{Example of a subsequence of periodic event times exhibiting fixed phase polling with period $P$.}
\label{fig:p1}
\end{figure}

Figure~\ref{fig:p1} shows an example of a possible subsequence of event data exhibiting fixed phase polling, where an error in one event does not propagate into future events. A cross indicates an observed event whilst a square indicates a period with missing data.

To construct a subsequence exhibiting fixed phase polling. For the $i$th polling subsequence and the $j^{th}$ period of the $i$th subsequence, $i=1,2,\ldots$ and $j = 1,\ldots,n_i$, let $\theta_{i,j,1}<\ldots<\theta_{i,j,m_{i,j}}$
be the order statistics from $m_{i,j}$ independent draws from the von Mises distribution $M(\pi,\kappa)$. 
Then for $k=1\ldots,m_{i,j}$ we define the ordered event times for that period to be
\begin{equation}
\label{eq:model1}
y'_{i,j,k}=\sum_{i'=1}^{i-1}(x_{i'}+n_{i'}P)+x_i+P\left\{(j-1)+\frac{\theta_{i,j,k}}{2\pi}\right\}.
\end{equation}
Considering these event times transformed to the unit circle, let
\begin{equation}
\label{AD1}
\phi_{i,j,k}=\frac{2\pi \cdot y'_{i,j,k} }{P}\pmod{2\pi},
\end{equation}
be the corresponding angular representation. Since $\theta_{i,j,k}$ has expected value $\pi$, it can easily be seen that the expected angular position of each event time in the $i$th periodic subsequence is given by
\begin{equation}
  \nu_i=\mathrm{E}(\phi_{i,j,k})= \left(\frac{2\pi\sum_{i'=1}^{i}x_{i'}}{P} +\pi\right) \pmod{2\pi}.
  \label{eq:nu_i}
\end{equation}
Therefore the angular displacement error associated with each event time is described by the $M(0,\kappa)$ distributed variables
\begin{align}
\label{AD2}
z_{i,j,k}&=(\phi_{i,j,k}-\nu_{i}) \pmod{2\pi}\\
&=(\theta_{i,j,k}+\pi)\pmod{2\pi}.
\end{align}

Finally, to complete the specification for a point process of event times with periodic subsequences, let $y_1<y_2<\ldots$ be the sequence of observable event times defined by
\begin{align}
  y_{\sigma(i,j,k)}&=y'_{i,j,k},\\
  \sigma(i,j,k)&=\sum_{i'=1}^{i-1}\sum_{j'=1}^{n_{i'}}m_{i',j'}+\sum_{j'=1}^{j-1}m_{i,j'}+k.
  \label{eq:data_indexing}
\end{align}
An illustration of this indexing format is provided in Figure \ref{fig:p1}.

\subsection{Fixed Duration Polling}
\begin{figure}[th]
  \centering
\includegraphics[width= \textwidth]{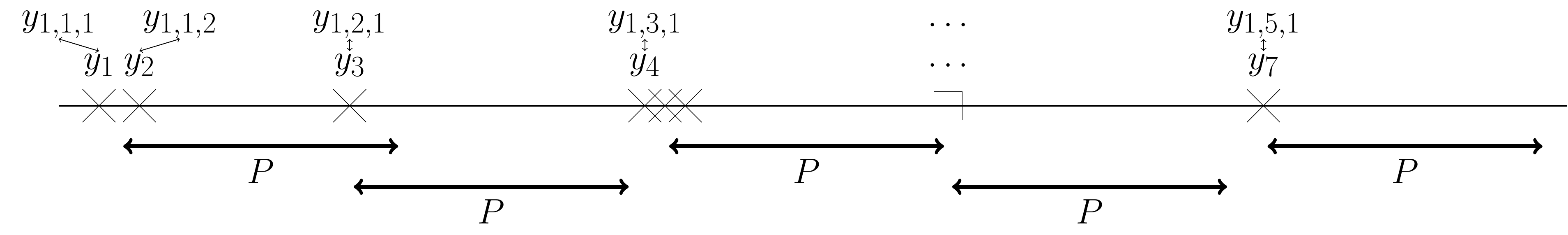} 
\caption{Example of a subsequence of periodic event times exhibiting fixed duration polling with period $P$.}
\label{fig:p2}
\end{figure}

An example of a possible subsequence exhibiting fixed duration polling is shown in Figure~\ref{fig:p2}. It is most intuitive to define the event times recursively, since the event times in one period affect the mean of the distribution of event times for subsequent periods.
To construct a subsequence of event times exhibiting fixed duration polling, let $z'_{i,j,k}\in(-\pi,\pi]$ be the angular displacement error associated with each event time, drawn from $M(0,\kappa)$. 
Let 
\begin{align}
\bar{y}_{i,j}=
\begin{cases}
x_1 & m_{i,j}=0,i=1,j=1,\\
\bar{y}_{i-1,n_{i-1}} +x_i & m_{i,j}=0,i>1,j=1,\\
\bar{y}_{i,n_{i-1}} +P & m_{i,j}=0,j>1,\\
\sum_{k=1}^{m_{i,j}}y'_{i,j,k}/m_{i,j}& m_{i,j}>0,
\end{cases}
\end{align}
 such that for each non-empty period in a beaconing subsequence $\bar{y}_{i,j}$ is the arithmetic mean of all event times in that period. Then we define the ordered event times for each period to be
\begin{align}
y'_{i,j,k}&= 
\begin{cases}
x_1+z'_{1,1,k}& i=1,j=1,\\
\bar{y}_{i-1,n_{i-1}}+x_i+z'_{i,1,k}  &i>1,j=1,\\
 \bar{y}_{i,j-1}+P+z'_{i,j,k}  &j>1.
 \end{cases}
\end{align}
The sequence of  observable event times is again described by \eqref{eq:data_indexing}.

\section{Changepoint Detection}
\label{sec:CD}
Changepoint detection techniques are widely used in data analysis across a range of scientific fields. These methods partition a sequence of data into a possibly unknown number of smaller segments, such that the data within each segment are assumed to arise from a single generative model. Here, discrete changepoint analysis are used to separate a point process of event times on a computer network edge into periodic subsequences, using one of the proposed models from Section \ref{sec:M}. 

Let $y_1<\ldots<y_n$ be a sequence of event times from a point process. Suppose the sequence is partitioned into $m+1$ segments by $m$ integer-valued changepoints $\bar{\tau}=(\tau_1,\ldots,\tau_m)$, ordered such that $0\equiv\tau_0<\tau_1<\ldots<\tau_m<\tau_{m+1}\equiv n$. For each $i=1,\ldots,m+1$, the $i$th segment of data is the subsequence of event times $y_{\tau_{i-1}+1:\tau_i}=(y_{\tau_{i-1}+1},\ldots,y_{\tau_i})$.

A common aim of discrete changepoint detection algorithms is to find changepoints that minimize an overall cost function
\begin{align}
\label{eq:3}
\sum_{i=1}^{m+1}[C(y_{\tau_{i-1}+1:\tau_i})]+\beta_n,
\end{align}
where $C$ is a segment-based cost function relating to the fitted likelihood of the data in a segment and $\beta_n\in\mathbb{R}$ is a penalty term to discourage over fitting.

There are many changepoint detection algorithms, but here we consider binary segmentation (BS), optimal partitioning (OP) and the pruned exact linear time (PELT) algorithm \cite{PELT}. \cite{scott} introduces the BS method for changepoint detection. This method has the advantage of being computationally efficient, $\mathcal{O}(n \log n)$, but it is not guaranteed to find the global minimum of \eqref{eq:3}.
\cite{jackson} introduces the OP method which is guaranteed to minimize \eqref{eq:3}. 
The OP method works by iterating sequentially through each event time and minimizing \eqref{eq:3} conditional on all previous combinations of changepoints.
The disadvantage to this method is that it has computational cost which is quadratic in $n$.

\label{sec:8}
\cite{PELT} introduces the PELT method which uses pruning to improve the computational efficiency of the OP method whilst still ensuring that the search algorithm finds a global minimum to \eqref{eq:3}. Under the assumption that the number of changepoints $m$ increases linearly with the size of the data $n$, PELT has a linear computational cost. Pseudo code for both the PELT and the OP methods can be found in \cite{PELT}.

Following the changepoint literature \cite{chen}, $C$ is chosen to be twice the negative log-likelihood function for a periodic subsequence of data according to a model from Section \ref{sec:M}. Furthermore, in accordance with the Bayesian information criterion (BIC) \cite{BIC}, the penalty is chosen to be of the form
\begin{align}
\label{eq:beta1}
\beta_n=\alpha\log n,
\end{align}
where $\alpha$ notionally represents the number of additional free parameters introduced to the model by adding a changepoint. Here the natural choice is $\alpha=2$, since the only parameters introduced by adding a changepoint are the position of the changepoint within the data sequence and the location parameter \eqref{eq:nu_i} of the new periodic subsequence.

\label{sec:CDPS}
Using the various definitions from Section \ref{sec:M}, the likelihood of a proposed subsequence is given by
\begin{equation}
\label{eq:cost}
\begin{split}
&\lambda \mbox{e}^{-\lambda x_i}
(1-q)^{n_i-1}q
\prod_{j=1}^{n_i}\left( (1-\delta_{i,j})p+\delta_{i,j}(1-p)(1-r)r^{m_{i,j}-1}
\prod_{k=1}^{m_{i,j}}\frac{\exp\{\kappa\cos(\phi_{i,j,k}-\hat{\nu}_{i,j})\}}{2\pi I_0(\kappa)}\right),
\end{split}
\end{equation}
where in the case of fixed phase polling $\hat{\nu}_{i,j}= \hat{\nu}_i$ is the MLE \eqref{nu1} for the location parameter of the von Mises distribution using the angular representation \eqref{AD1} of all event times from the proposed subsequence. In the case of fixed duration polling $\hat{\nu}_{i,j}$ is the MLE \eqref{nu1} for the location parameter of the von Mises distribution using the angular representation \eqref{AD1} of all event times from the previous non-empty period.


When the parameters $(p,q,r,\lambda,\kappa)$ are considered unknown, independent conjugate priors can be deployed such that $p$, $q$ and $r$ have beta distributions with respective parameters $(\alpha_p,\beta_p)$, $(\alpha_q,\beta_q)$, and $(\alpha_r,\beta_r)$, whilst $\lambda\sim \mbox{gamma}(\alpha_{\lambda},\beta_{\lambda}).$ The prior distribution for $\kappa$ is given by \eqref{VMprior}.
A simple Metropolis Hastings algorithm with a uniform proposal density centered at the current parameter value can be used to obtain a posterior estimate of $\kappa$, \eqref{VMpos} using the estimated angular displacements of the event times, 
$$\hat{z}_{i,j,k}=(\phi_{i,j,k}-\hat{\nu}_{i,j}) \pmod{2\pi},$$
where it is assumed that $\hat{z}_{i,j,k}\sim M (0,\kappa)$.

The proposed procedure is to iterate between three inferential steps:
\begin{enumerate}
\item Changepoint analysis for identifying periodic subsequence using PELT\label{item:cp}
\item Bayesian estimation of the nuisance parameters  $(p,q,r,\lambda,\kappa)$ \label{item:bayes}
\item Updating the estimated periodicity $P$ \label{item:period}
\end{enumerate}
The algorithm simply loops through steps \ref{item:cp}-\ref{item:period} repeatedly until convergence is reached: Convergence is determined to have occurred when identical changepoints are found by step \ref{item:cp} in two successive iterations of the algorithm. After each iteration of the changepoint algorithm in step \ref{item:cp}, the nuisance parameters are re-estimated using their revised posterior means, conditional on the updated set of changepoints, in step \ref{item:bayes}. The posterior means of the parameters $p,q,r,\lambda$ all have closed form under the conjugate priors, and the posterior mean of $\kappa$ is estimated by Metropolis Hastings sampling as described above. Finally, the estimated periodicity $P$ is also updated in step \ref{item:period} as follows: For each non-empty period in a beaconing subsequence, let $\bar{y}_{i,j}$ be the arithmetic mean of all event times in that period; then for $j=1,\ldots,n_i-1$ let $w_{i,j}=\bar{y}_{i,j+1}-\bar{y}_{i,j}$ be the difference between the mean event times from two successive periods of events. To reduce the influence of missing event data, $P$ is estimated to be the median value of $\{w_{i,j}\}$ over all subsequences $i$ defined by the current set of changepoints.

Since computational scalability is paramount, changepoints are detected using PELT, which has a computational cost which is linear in the number of data points. For updating model parameters between successive iterations of the PELT algorithm, maximum likelihood estimation can potentially lead to degenerate solutions and therefore a Bayesian estimation procedure is adopted.


\section{Examples}
\label{sec:7}
\subsection{Simulated Data Example}

To gain an understanding of the accuracy and the efficiency of two of the changepoint detection algorithms, both the BS and PELT methods described in Section \ref{sec:CD} are applied to sequences of events times sampled from the full generative model for fixed phase polling from Section \ref{sec:M}. The aim is to partition the event times so that each changepoint is positioned at the start of a new periodic subsequence, therefore changepoints can only occur at the discrete number of event times within the data.

Three different combinations of the model parameters $p,q,r,\lambda$ and $\kappa$ are used to generate different types of periodic subsequences which could arise in real world computer network data. For example, larger values of $\kappa$ induce smaller error terms in the angular displacements of the beaconing subsequences, whilst increasing $\lambda$ leads to longer lengths of inactivity between the subsequences. Both of these changes should make separate periodic subsequences easier to identify. Without loss of generality, the periodicity is set to $1$. The three sets of parameter choices are presented in Table \ref{table:parameters}. For each set of parameters, $100$ sequences of $10$ beaconing subsequences were generated. 

\begin{table}[h!]
  \begin{center}
    \pgfkeys{/pgf/number format/fixed}
    \begin{filecontents*}{simulation_parameters.txt}
9,10,0.05,0.5,0.1,0.05
8,10,0.1,0.5,0.1,0.1
7,10,0.2,0.5,0.1,0.2
\end{filecontents*}
\pgfplotstabletypeset[col sep=comma,
      every head row/.style={before row=\toprule,after row=\bottomrule},
      every last row/.style={after row=\bottomrule},
      display columns/0/.style={column name=Parameter settings},
      display columns/1/.style={column name=$\kappa$,},
      display columns/2/.style={column name=$p$},
      display columns/3/.style={column name=$r$},
      display columns/4/.style={column name=$q$},
      display columns/5/.style={column name=$\lambda$},
      create on use/new col/.style={create col/set list={1,...,3}},
      columns={new col,0,1,2,3,4}
    ]{simulation_parameters.txt}
\end{center}
\caption{Parameter settings for generating periodic subsequences of event times from the generative model for fixed phase polling in Section \ref{sec:M}.}
\label{table:parameters}
\end{table}

Both the BS and the PELT algorithms identify changepoints by
minimizing \eqref{eq:3}.
In this example, the proportion of true positive(TP) changepoints corresponds to the number of true changepoints correctly identified divided by the total number of true changepoints, whilst the proportion of false positive(FP) changepoints corresponds to the number of events incorrectly identified as changepoints, divided by the true number of events which do not correspond to a changepoint.

\begin{figure}[ht!]
  \centering
    \includegraphics[width=\textwidth]{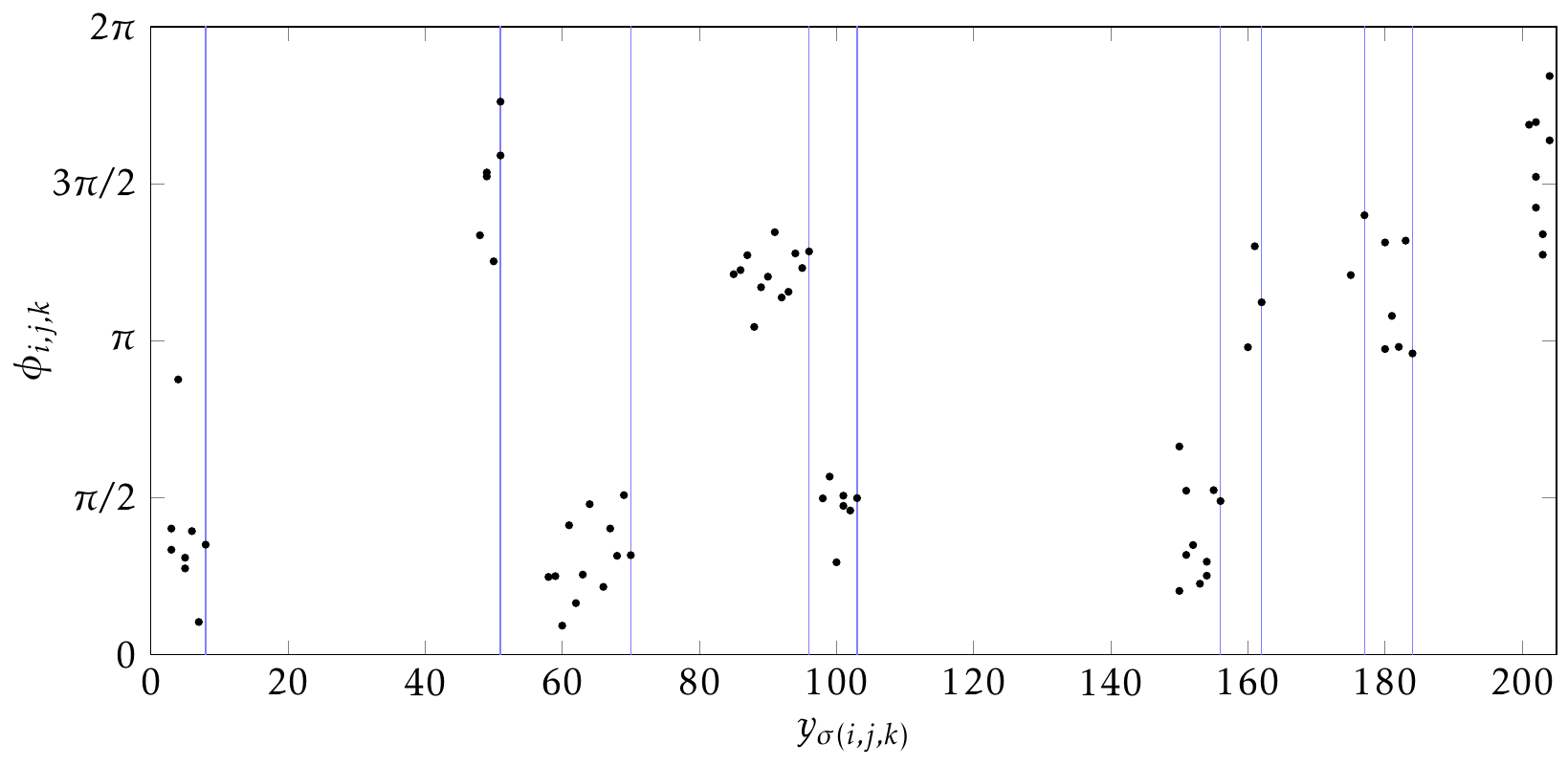}
\caption{A simulated point process representing periodic subsequences, partitioned by changepoints  found using PELT.}
\label{fig:7}
\end{figure}

Figure \ref{fig:7} plots an example of one sequence of simulated event times comprised of ten beaconing subsequences with periodicity $1$, generated with parameter setting $2$ from Table \ref{table:parameters}. Changepoints were detected using PELT with the commonly used BIC penalty $\beta_n=2\log(n)$ \eqref{eq:beta1}. For each event time $y_{\sigma(i,j,k)}$ from \eqref{eq:data_indexing}, the angular position $\phi_{i,j,k}$ from \eqref{AD1}, determined by the fitted changepoints, is plotted. The vertical lines indicate the locations of the changepoints, which correctly partition the event times into the ten separate subsequences.

 To obtain a thorough comparison of the two
methods, the performance of both methods was tested over a range of settings for the penalty $\beta_n$ from \eqref{eq:beta1} by repeating the analysis for
$\alpha\in\{-3,-1.5,0,2,5,10,20,40\}$. 
Increasing $\alpha$ corresponds to a higher penalty $\beta_n$ for adding a changepoint, therefore we would expect to detect fewer TP and FP changepoints as $\alpha$ increases. 

\begin{figure}[h]
  \includegraphics[width=\textwidth]{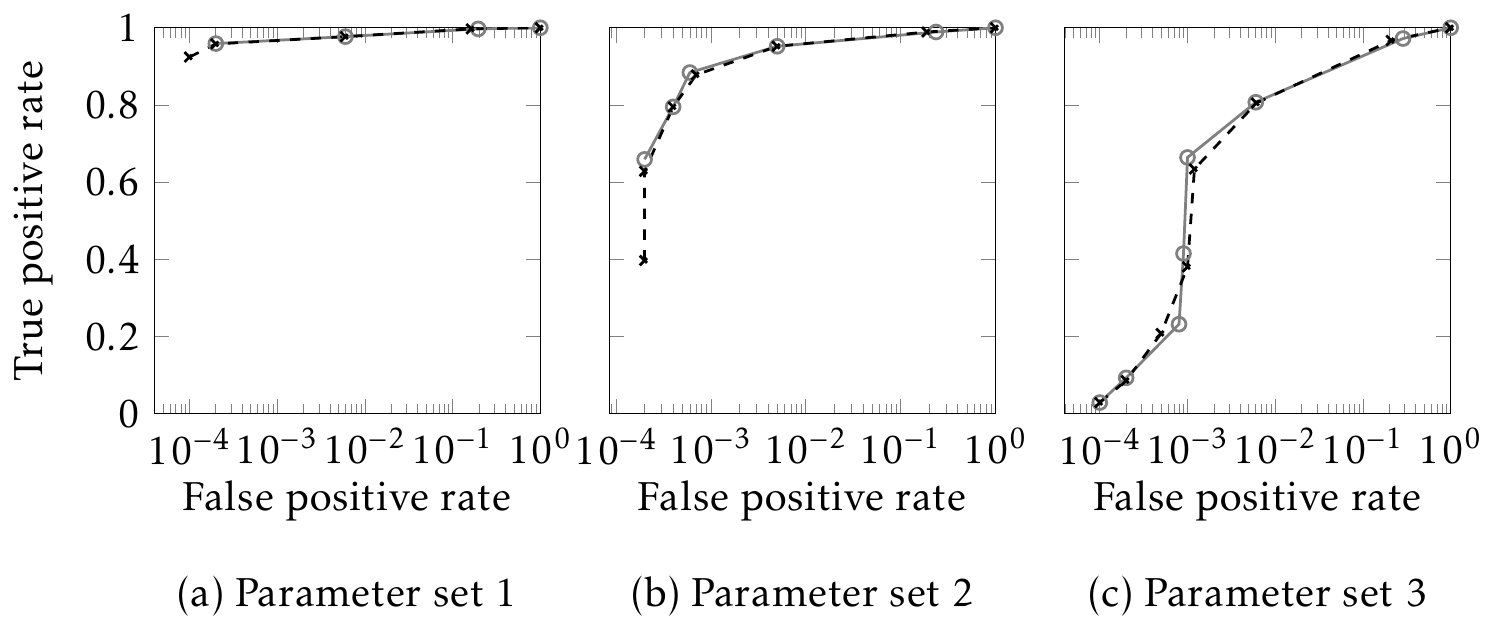}
\caption{ROC curves comparing the changepoints detected using PELT and BS for data generated using the three parameter settings in Table \ref{table:parameters}.}
\label{figure:A1}
\end{figure}

Note that for completeness even negative penalties are considered, in order to reach the extreme
case where every event time is selected to be a changepoint. This
enables a receiver operating characteristic (ROC) curve to be plotted
in Figure \ref{figure:A1}, showing how the true and false positive
changepoint rates increase as the penalty term is reduced. If $\alpha$ was increased further, we would eventually reach the other extreme where no event times are selected as changepoints. Since PELT only has linear computational cost when the number of changepoints increases linearly with the size of the data, \cite{PELT} reducing the number of changepoints towards zero is computationally prohibitive and is therefore not included. 

The proportions of true and false positive changepoints found using PELT
closely resemble those found using the BS method. Table \ref{table:A}
presents the proportion of TP and FP changepoints found for the
commonly used BIC penalty, $\beta_n=2\log(n)$. In this case PELT
identifies at least as high a proportion of TP changepoints as the BS
method, and at least as small a proportion of FP changepoints for all
three parameter settings.

\begin{table}[h!]
\begin{center}
\begin{tabular}[h!]{c|c|c|c|c}
\multirow{2}{*}{Parameter settings}&\multicolumn{2}{c|}{BS}&\multicolumn{2}{c}{PELT}\\
\cline{2-5}
&TP&FP&TP&FP\\
\hline
$1$&$0.9587$&$0.0002$&$0.9587$&$0.0002$\\
$2$&$0.8784$&$0.0007$&$0.8845$&$0.0006$\\
$3$&$0.6340$&$0.0012$&$0.6639$&$0.0010$\\
\end{tabular}
\end{center}
\caption{Proportion of true positive (TP) and false positive (FP) changepoints identified under the BIC penalty.}
\label{table:A}
\end{table}

To further investigate how the two methods differ, the run time of the two algorithms is measured against the number of changepoints for the commonly used BIC penalty. For each set of parameters from Table \ref{table:parameters}, ten sequences formed of different numbers of beaconing subsequences (equivalently, changepoints) were generated. The average run time in seconds on a computer with an Intel Core i7 processor, clocked at 2.2 GHz, are presented in Table \ref{table:times}. Both algorithms use the same underlying code for calculating the cost of a proposed changepoint configuration. In almost all situations, the PELT method was quicker than the BS method, especially when the number of changepoints in the model increased.

\begin{table}[h!]
\begin{center}
\begin{tabular}[h!]{c|c|ccc|ccc}
\multicolumn{2}{c|}{Changepoint algorithm}&\multicolumn{3}{c|}{BS}&\multicolumn{3}{c}{PELT}\\
\hline
\multicolumn{2}{c|}{Parameter setting}&1&2&3&1&2&3\\
\hline
\multirow{4}{*}{Number of changepoints}&$10$&$14.1$&$8.7$&$8.2$&$12.1$&$6.2$&$9.3$\\
&$20$&$46.8$&$36.4$&$29.0$&$28.3$&$19.9$&$20.4$\\
&$50$&$293.0$&$219.4$&$171.3$&$57.7$&$45.0$&$63.4$\\
&$100$&$1110.1$&$995.9$&$824.7$&$102.9$&$99.7$&$200.4$\\
\end{tabular}
\end{center}
\caption{Average run times (in seconds) using the two changepoint detection algorithms for different numbers of changepoints and parameter settings from Table \ref{table:parameters}.}
\label{table:times}
\end{table}

When monitoring real computer networks, traffic is often observed for several days to allow construction of an accurate model of normal behavior. It is therefore possible that a large number of beaconing periodic subsequences could be observed; in such cases, where there is an unknown but potentially large number of changepoints, the PELT method is preferable since it has a linear computational cost.

\subsection{Real Data Example}
\label{sec:RD}
The PELT changepoint detection algorithm for detecting periodic subsequences, presented in Section \ref{sec:CD}, is applied to the authentication data from the Los Alamos National Laboratory (LANL) computer network described in Section \ref{sec:CN}. The analysis in this section focuses on the ``Log On'' events for user U$514$ depicted in Figure \ref{fig:logon}. The event times for each source-destination pair of computers are modeled as separate point processes. The time of day distribution of all Log On events from user U$514$ are shown in Figure \ref{fig:logon}, where the lack of variation in the number of events within and outside the working day indicated the presence of substantial polling behavior associated with this username. The periodicity detection method was applied separately to each computer-computer point process, identifying the first event after each changepoint as a user-driven Log On event. Unlike the previous example these data exhibit fixed duration polling, where any error in the beaconing subsequence propagates into future event times (see Section $\ref{sec:M}$).

To initialize the algorithm, the model parameters $(p,q,r,\lambda,\kappa)$ are initially estimated using the means of the prior distributions given in Section \ref{sec:CDPS}. The hyper-parameters of the prior distributions are estimated empirically from a small segment of a beaconing subsequence, in one point process in the network. These hyper-parameters are given in Table \ref{table:hyperparameters1}.

\begin{table}[h!]
  \begin{center}
    \pgfkeys{/pgf/number format/fixed}
    \begin{filecontents*}{simulation_parameters_1.txt}
9,10,10,1,10,5,10,1,20,1,

\end{filecontents*}
\pgfplotstabletypeset[col sep=comma,
      every head row/.style={before row=\toprule,after row=\bottomrule},
      every last row/.style={after row=\bottomrule},
      display columns/0/.style={column name=Parameter settings},
      display columns/1/.style={column name=$R$,},
      display columns/2/.style={column name=$c$},
      display columns/3/.style={column name=$\alpha_p$},
      display columns/4/.style={column name=$\beta_p$},
      display columns/5/.style={column name=$\alpha_r$},
      display columns/6/.style={column name=$\beta_r$},
      display columns/7/.style={column name=$\alpha_q$},
      display columns/8/.style={column name=$\beta_q$},
      display columns/9/.style={column name=$\alpha_{\lambda}$},
      display columns/10/.style={column name=$\beta_{\lambda}$},
      create on use/new col/.style={create col/set list={1}},
      columns={new col,0,1,2,3,4,5,6,7,8,9}
    ]{simulation_parameters_1.txt}
\end{center}
\caption{Parameter settings for generating periodic subsequences of event times from the generative model in Section \ref{sec:M}.}
\label{table:hyperparameters1}
\end{table}

\begin{figure}[!ht]
\begin{minipage}[c]{0.5\textwidth}
  \centering
  \includegraphics[width=\textwidth]{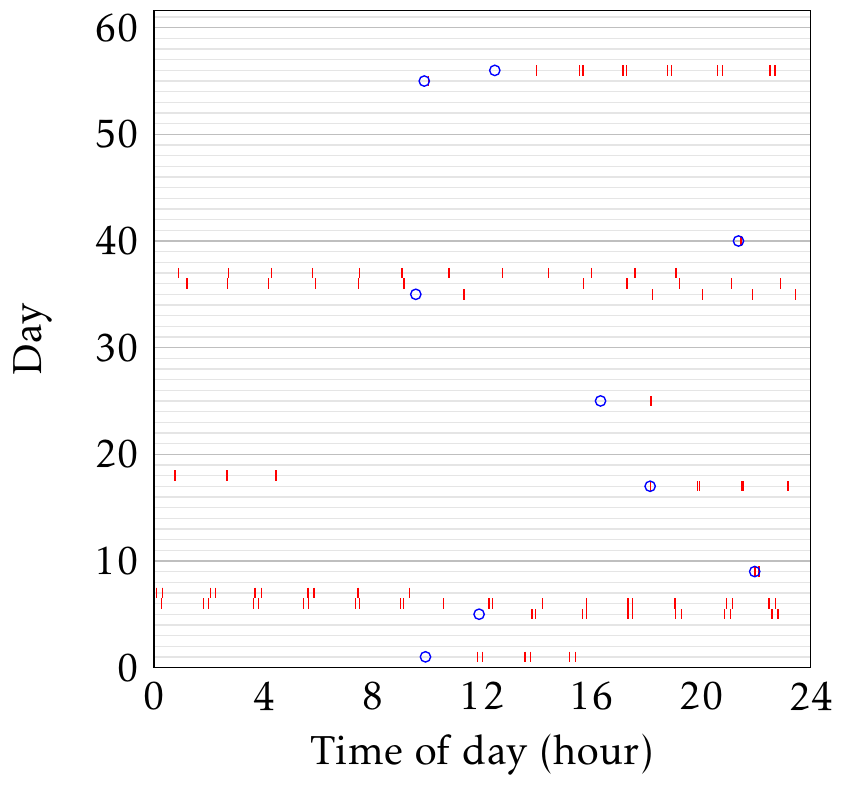}
\end{minipage}
\hfill
\begin{minipage}[c]{0.5\textwidth}
\centering
  \includegraphics[width=\textwidth]{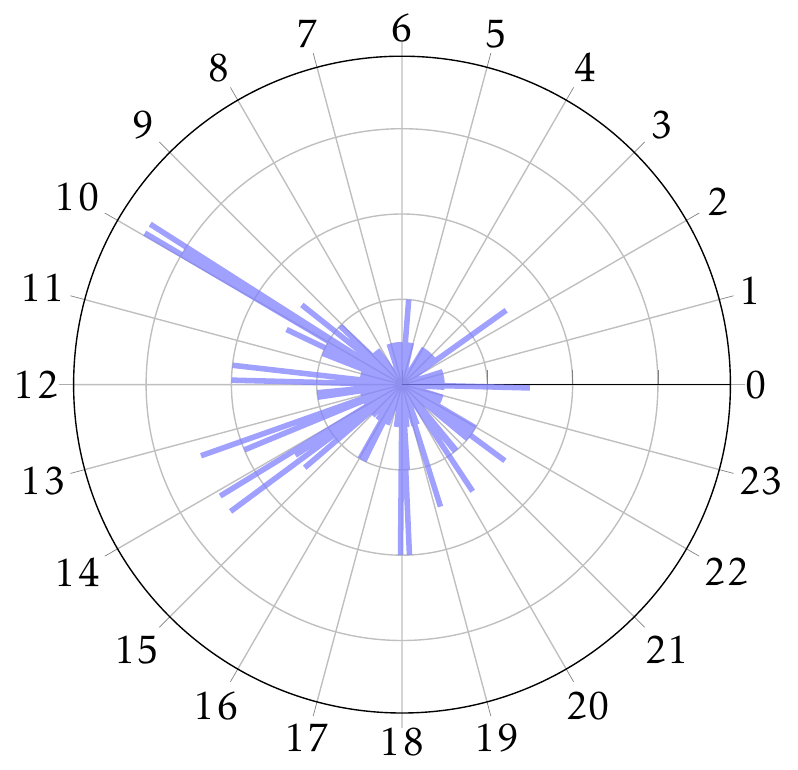}
\end{minipage}
\caption{Log on event times for U514 from the LANL computer network. Left: Log on event times between the two specific computers C$528$ and C$15607$, where events have been attributed as human or automated, periodic events. Right: The time of day distribution for all human-generated events attributed to the user.}
\label{fig:logon2}
\end{figure}

As in Figure \ref{fig:logon}, the left panel of Figure \ref{fig:logon2} plots all event times over the entire 58-day period, where user U$514$ logs onto computer C$528$ from computer C$15607$. The circles in the diagram now indicate the inferred user-driven events that initiate the start of an automated polling subsequence. It is clear that meaningful subsequences of periodic event times are identified, and the methodology is robust to duplicate event data, as seen between days 5 and 10, and missing event data, as seen between days 35 and 40.

The right panel of Figure \ref{fig:logon2} plots the distribution of the $169$ human Log On events found by applying the PELT changepoint detection algorithm to all pairs of computers used by user U$514$ over the entire 58-day period. The distribution of these events is much more consistent with human behavior than Figure \ref{fig:logon}, with the majority of events occurring within the hours of an extended working day (9am-8pm) and a large spike of Log On events at the start of the working day (9am-10am).

For user U$514$, automated polling behavior accounts for over $97\%$ of all Log On data. As a data reduction tool, this represents a significant thinning of the bulk of authentication data. Further analytics operating on only user-driven behavior would become more manageable and allow for more robust modeling, potentially strengthening anomaly detection capabilities.

\subsubsection{Robustness}

The proposed procedure of iterating between the changepoint analysis and Bayesian estimation of the nuisance parameters introduces random variation to the algorithm due to the stochastic nature of the Metropolis-Hastings (MH) algorithm. Variation can also be introduced by varying the specification of the prior hyper-parameters displayed in Table \ref{table:hyperparameters1}. We examine the robustness of the model by performing $10$ repetitions of the algorithm on the same data set, for each of three different sets of prior hyper-parameters. These hyper-parameters are presented in Table \ref{table:hyperparameters2}.
\begin{table}[h!]
  \begin{center}
    \pgfkeys{/pgf/number format/fixed}
    \begin{filecontents*}{simulation_parameters_2.txt}
8,10,10,2,10,5,10,1,10,1,
9,10,10,1,10,5,10,1,20,1,
7,10,10,3,10,5,10,1,5,1,
\end{filecontents*}
\pgfplotstabletypeset[col sep=comma,
      every head row/.style={before row=\toprule,after row=\bottomrule},
      every last row/.style={after row=\bottomrule},
      display columns/0/.style={column name=Parameter settings},
      display columns/1/.style={column name=$R$,},
      display columns/2/.style={column name=$c$},
      display columns/3/.style={column name=$\alpha_p$},
      display columns/4/.style={column name=$\beta_p$},
      display columns/5/.style={column name=$\alpha_r$},
      display columns/6/.style={column name=$\beta_r$},
      display columns/7/.style={column name=$\alpha_q$},
      display columns/8/.style={column name=$\beta_q$},
      display columns/9/.style={column name=$\alpha_{\lambda}$},
      display columns/10/.style={column name=$\beta_{\lambda}$},
      create on use/new col/.style={create col/set list={1,2,3}},
      columns={new col,0,1,2,3,4,5,6,7,8,9}
    ]{simulation_parameters_2.txt}
\end{center}
\caption{Parameter settings for generating periodic subsequences of event times from the generative model in Section \ref{sec:M}.}
\label{table:hyperparameters2}
\end{table}

To measure the effect of the random variation introduced by the MH algorithm for approximating Bayesian estimation of the posterior parameters, let $e_1,\ldots, e_M$ be the ordered list of all events in the data sequence. Then let 
$
\label{binary}
b_{i,l,1},\ldots, b_{i,l,M}
$
 be a binary representation of the changepoint events found by the $i$th repetition of the algorithm, with prior hyper-parameters $l\in \{1,2,3\}$. When the algorithm identifies $e_k$ to be a changepoint (user driven event), $b_{i,l,k} = 1.$ Otherwise $e_k$ is assumed to be automated and $b_{i,l,k}=0$. 
For prior hyper-parameter settings $l\in\{1,2,3\}$ and each pair of repetitions of the algorithm $i,j \in \{1\ldots 10\},$ $i\neq j$, let 
$$ c_{i,j,l} = \frac {\sum_{k=1}^M \mid b_{i,l,k} - b_{j,l,k} \mid }{M}$$
be the proportion of events which  are assigned differently. $c_{i,j,l}$ is calculated for all pairs of repetitions of the algorithm under each choice of hyper-parameters.
Finally let $c_l$ 
be the mean of $c_{i,j,l}$ over all pairs of repetitions, $i,j\in1\ldots 10,$ $i\neq j$.

 For the first set of hyper-parameters, we find that $c_1=0.000103.$ This value implies that any two repetitions of the algorithm make identical inference except for approximately $1$ in every $1000$ events differently. We also found that $c_2=0.000177$ and $c_3= 0.000103$.

We are also interested in finding the effects of changing the prior hyper-parameterisation of the model. 
For each prior hyper-parameterisation displayed in Table \ref{table:hyperparameters2} and event $e_k$, let 
$$\bar{b}_{l,k} =\frac{ \sum_{i=1}^{10} b_{i,l,k}} {10} $$
be the proportion of repetitions of the algorithm which identify the $k^{th}$ event to be user-driven. 
Then, for $l_1,l_2\in\{1,2,3\}$, let
$$ d_{l_1,l_2} = \frac {\sum_{k=1}^M \mid \bar{b}_{l_1,k} - \bar{b}_{l_2,k} \mid }{M}$$
be the average proportion of events which are differently categorized for prior hyper-parameterisations $l_1$ and $l_2$.  

For the $2$ alternative sets of hyper-parameters proposed in Table \ref{table:hyperparameters1} we find that
 \begin{align}
 d_{1,2}=0.000332,\quad
 d_{1,3}=0.000442.
 \end{align} 
 Compared to the results found for the original parameter settings, approximately three in every $1000$ events are assigned differently when using prior hyper-parameter settings $2$, and
 approximately four in every $1000$ events are assigned differently when using prior hyper-parameter settings $3.$
 
The example presented in the left panel of Figure \ref{fig:logon2} identified the inferred user-driven Log-On events when U$514$ logs onto computer C$528$ from computer C$15607$.
In this example, the algorithm finds the same inferred user-driven Log-On events for all repetitions of the algorithm, for all prior hyper-parameterisations. 
 The right panel of Figure \ref{fig:logon2} plots the distribution of the $169$ human Log On events found by applying the algorithm to all pairs of computers used by user U$514$ over the entire 58-day period. This plot is visually indistinguishable for all repetitions of the algorithm, for all prior hyper-parameterisations.
 
 The results in this section show that any variability in the algorithm, due to the stochastic variability of the MH algorithm or from altering the specification of the prior hyper-parameters, has a very small effect on the resulting inference.

\section{Conclusion}
In this article we first demonstrated that intermittent polling behavior in computer network data can typically be detected using standard Fourier analysis methods, such as \cite{polling}. Secondly, we proposed a methodology for identifying the separate beaconing subsequences within intermittent polling data.
Conditional on the presence of intermittent polling behavior, with a constant periodicity estimated by \cite{polling}, we introduced a changepoint detection methodology to identify the separate polling subsequences within the overall event sequence. We proposed iterating between the changepoint analysis and estimation of nuisance parameters, to reflect the dependency between the these quantities.  The proposed method is robust to complications typically encountered in computer event data collection, such as duplicated and missing data. 

The purpose of partitioning event data into periodic subsequences is to identify potentially user-driven actions in computer network event data, with the future intention of modeling user-driven and automated behavior separately for anomaly detection purposes. User-driven actions are identified as the first events of each beaconing subsequence, which are followed by automated periodic updates.

Treating automated and user-driven data separately should provide a more robust framework for modeling computer network data, whereby bespoke models can be specified for each type of behavior (see for example \cite{neil}). Alternatively, as a data reduction tool, identifying only user-driven events represents a significant thinning of the bulk of computer network data, potentially improving anomaly detection capability and efficiency.

In this article the changepoint detection methodology was applied separately to each edge in a computer network, with the aim of classifying human user activity against automated events.  
If the events from an edge in the network consist of 
a single beaconing sequence  with no phase shifts, this suggests no human activity and the event data will be best described using the simpler method of \cite{polling}. However, if the events consist of multiple, intermittent periodic subsequences sharing a constant periodicity, the methodology proposed in this article can be used to identify the user-driven events initiating each period subsequence. If the events exhibit no periodic patterns, then we might conclude that the events from that edge are potentially all user-driven.

The algorithm was applied to authentication data from the Los Alamos National Laboratory (LANL) enterprise computer network. The algorithm identified meaningful subsequences of periodic event times and was robust to duplicate and missing event data. In the absence of true classification labels for the data, we investigated the time of day distributions of the inferred user-driven events and compared this with the distribution for all of the authentication events from one user. The distribution of the inferred user-driven events was much more consistent with human behavior, displaying a more clear diurnal pattern. 

Further extensions of the model could be considered. For example, within the $i$th subsequence, $\theta_{ijk}$ are assumed to be the order statistics of independent identically distributed draws from a $M(\pi,\kappa)$ distribution. However, in reality there is often a correlation exhibited between these angular displacements within a beaconing period, as events can arrive in bursts. To incorporate this correlation into the model, we could postulate a hierarchical model for beaconing subsequences of event times, where clusters of event time variables $\theta_{ijk}$ are considered to be only conditionally independent given some unobserved location value.

\bibliographystyle{vancouver}
\bibliography{refs}

\end{document}